\documentclass[12pt]{article}
\usepackage{setspace}
\usepackage[a4paper, top=35mm, bottom=35mm, left=35mm, right=35mm]{geometry}
\usepackage{siunitx}
\usepackage{graphicx}
\usepackage{amsmath}
\usepackage[T1]{fontenc}
\usepackage{newpxtext,newpxmath}
\usepackage{natbib}
\usepackage[dvipsnames]{xcolor}
\usepackage{hyperref}
\usepackage{placeins}
\usepackage{standalone}
\usepackage{fancyhdr}
\usepackage{titlesec}
\usepackage{xspace}
\setcitestyle{authoryear,open={(},close={)}}
\hypersetup{
	colorlinks=true,
	linkcolor=Mulberry,
	citecolor=Mulberry,
	urlcolor=blue
}

\begin{document}

\title{Tracing Outflows from Stellar Feedback in the Early Universe with Lyman-$ \alpha $}

\author{James Nianias$ ^1 $, Jeremy Lim$ ^1 $, Yik Lok Wong$ ^2 $, Gordon Wong$ ^1 $,\\
	 Ishika Kaur$ ^1 $, and Wenjun Chen}
\date{%
	\small{
	$ ^1 $Department of Physics, University of Hong Kong, Pokfulam Road, Hong Kong\\%
	$ ^2 $Department of Physics and Astronomy, The University of Manchester, Oxford Road, Manchester M13 9PL, UK\\%
}
}


\maketitle

\begin{abstract}
	Blind spectroscopy of massive lensing galaxy clusters with MUSE has revealed large numbers of gravitationally-lensed Lyman-$ \alpha $ emitters exhibiting asymmetric profiles at $ 2.9 \leq z \leq 6.7 $, suggesting abundant outflows from low-mass star-forming galaxies in the early universe.  Are these primaeval galaxies experiencing their first bursts of star formation, or established galaxies experiencing rejuvenation?  With JWST rest-frame optical/NIR continuum imaging now available for many of these objects, we can search for older stellar populations.  Here, we search for spectroscopic confirmation of outflows from these galaxies, finding a few high-signal-to-noise cases in which blueshifted interstellar absorption lines are detected.  Next, we analyse the star formation histories with combined HST + JWST photometry.  We find most them to be well characterised by very young, low metallicity stellar populations.  However, despite the rest-frame optical/NIR coverage of JWST, we cannot place strict upper bounds on the mass in old stars (age $ > 100\,\text{Myr} $).
\end{abstract}

\section{Introduction}

	Since their first detections in the mid-1990s, Lyman-$ \alpha $ emitters (LAEs) have become important tools with which to study the early universe.  With specific star formation rates (sSFRs) on or above the main sequence and low dust content \citep{iani24}, much of the radiation from beyond the Lyman limit ($ \lambda < 912\,\AA $) is reprocessed into the Lyman-$ \alpha $ line, which consequently can have extremely large equivalent width (EW).  This makes LAEs easy to detect in narrowband imaging (e.g. \citealt{ouchi03}) as well as blind spectroscopy (e.g. \citealt{bacon17}) despite their generally low masses ($ 10^6 \lesssim M_{\star} /\text{M}_\odot  \lesssim 10^9 $; \citealt{iani24}).  Spectroscopic detection is also made easier by the fact that Lyman-$ \alpha $'s resonant nature couples it to the kinematics of the interstellar medium (ISM), producing a distinctive and easily identifiable asymmetric line profile.  Line profiles of this kind can be reproduced using expanding shells of gas with varying outflow velocities, neutral covering fractions, and dust content \citep{ahn04, verhamme06, orsi12}.  Stacked spectra of LAEs and Lyman break galaxies (LBGs) at high redshift, as well as some individual gravitationally-lensed galaxies, also exhibit blueshifted interstellar absorption features, cementing the link between asymmetric Lyman-$ \alpha $ profiles and outflows \citep{franx97, shapley03, berry12, jones12}.
	
	\cite{richard21} published spectro-photometric catalogues based on blind spectroscopy of massive lensing clusters with the Multi-Unit Spectroscopic Explorer (MUSE) and imaging with the Hubble Space Telescope (HST).  These catalogues cover a total of 12 lensing clusters, and include $ \sim 1000 $ lensed images of LAEs with redshifts from 2.9 to 6.7 (i.e. into the epoch of reionization).  The magnification provided by gravitational lensing (up to a factor of $ \sim 100 $) enables detections of intrinsically fainter LAEs than would otherwise be possible, and also permits measurements of the physical sizes of the UV continuum counterparts with HST down to the scale of just a few tens of pc \citep{claeyssens22} -- the size of individual globular clusters \citep{vanzella17}.  The nature of these faint and compact systems is still uncertain: are we seeing star forming regions in established galaxies, or primaeval star bursts in low-mass dark matter halos?  With the enhanced wavelength coverage, sensitivity, and spatial resolution afforded by the James Webb Space Telescope (JWST), we can begin to search for signs of older stellar populations in these objects by observing the rest-frame optical/near-IR (NIR).
	
	In Section \ref{section:spectra}, we use the MUSE spectra to confirm outflow signatures in these objects, comparing the Lyman-$ \alpha $, optically thin high-ionization nebular emission, and interstellar absorption lines.  To answer the question of whether these objects host older stellar populations, we perform SED fitting of the HST + JWST photometry, presented in Section \ref{section:sedfitting}.
	
\section{MUSE Spectroscopy}
\label{section:spectra}
	\begin{figure}[ht!]
	\centering
	  \includegraphics[width=\textwidth]{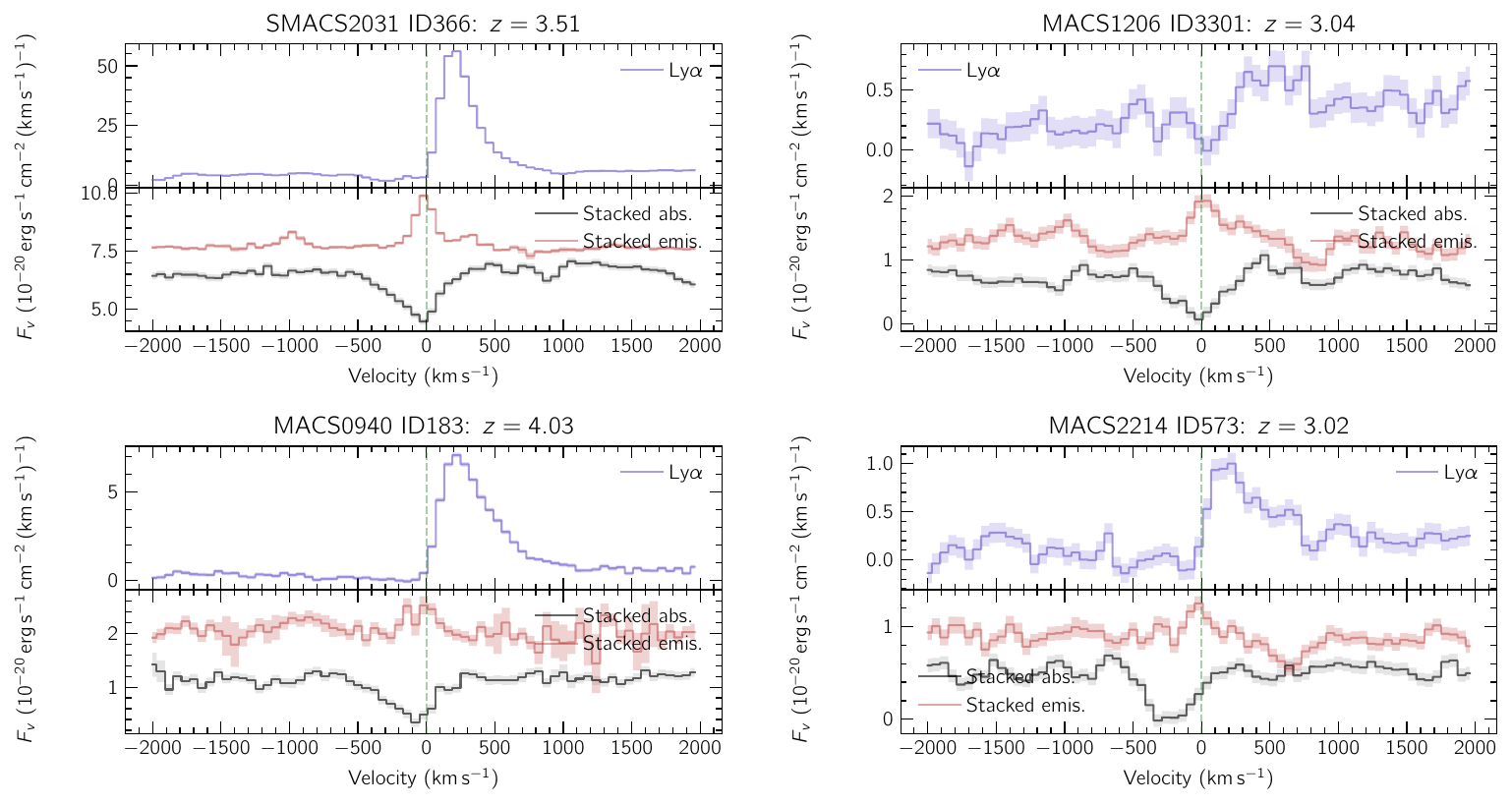}
	  \caption{Examples of MUSE spectra from Lyman-$ \alpha $ emitters where we find strong evidence for low-ionization interstellar absorption features.  The Lyman-$ \alpha $ profiles are shown in the upper panels in blue, and the stacked emission and absorption lines in the lower panels in red and black respectively.  The spectra show clear evidence for outflows in that the absorption lines appear either blueshiftee with respect to the emission, or else extend to more blueshifted velocities, while the Lyman-$ \alpha $ lines are asymmetric and redshifted relative to the systemic.}
	  \label{fig:lyaspecs}
	\end{figure}
	We extracted 1D spectra for each source from the corresponding MUSE cubes.  We chose only those sources with Lyman-$ \alpha $ detected with a signal-to-noise ratio (SNR) of 3 or better according to the catalogues of \cite{richard21}.  We then fitted analytic line profiles to each spectral line in listed in the catalogues: Lyman-$ \alpha $, nebular emission, and interstellar absorption lines.  For the nebular emission and interstellar absorption lines, we used gaussian profiles.  However, in order to model the asymmetric shape of the Lyman-$ \alpha $ line, we instead used an asymmetric gaussian profile as in \citep{shibuya14}.  Since Lyman-$ \alpha $ lines sometimes also have weaker secondary peaks blueshifted relative to the primary, we attempted to fit a blueshifted component as well.  In cases where the blueshifted peak (henceforth blue peak) was statistically significant and well-resolved from the primary (henceforth red peak), we considered it a detection.
	
	Having fitted all lines in each spectrum, We searched for spectra with kinematic signatures of outflows.  First, we selected those with statistically significant nebular emission lines and absorption lines.  The nebular emission lines are valuable as they allow us to measure the systemic redshift of the star-forming regions responsible for the Lyman-$ \alpha $ emission, which cannot be done using Lyman-$ \alpha $ itself due to the aforementioned asymmetric line profiles.  The interstellar absorption lines, on the other hand, can reveal outflowing gas screening the continuum source if they are blueshifted.  Confident identification of genuine emission and absorption features (besides Lyman-$ \alpha $) is complicated by the presence of residual atmospheric features in the MUSE spectrum.  However, we identified four confident examples of Lyman-$ \alpha $ emitters with both emission and absorption lines.  Comparing these lines in velocity space with the Lyman-$ \alpha $ profile, we find clear evidence of outflows: the absorption is blueshifted relative to the systemic redshift, or at least extends to more blueshifted velocities than the nebular emission, while the Lyman-$ \alpha $ is redshifted, with a skew towards higher velocities.  We show a comparison of the Lyman-$ \alpha $, emission, and absorption lines from two of these spectra in Figure \ref{fig:lyaspecs}.  These kinematics agree with those found in studies of other LAEs/LBGs (e.g. \citealt{shapley03,jones12,berry12}).
	
	All the spectra we find with unmistakable absorption features also have singly-peaked Lyman-$ \alpha $ emission (i.e. no blue peaks).  While the number of spectra is small, some of the Lyman-$ \alpha $ profiles have extremely high SNR (e.g. panels (a) and (b) of Figure \ref{fig:lyaspecs}), suggesting that this is not merely a consqeuence of blue peaks being below the noise level.  Qualitatively, this fits with the picture presented in \cite{erb15}, in which there is a greater covering fraction of neutral gas in single-peaked Lyman-$ \alpha $ spectra, which also gives rise to the absorption lines (which are mostly low-ionization species associated with neutral gas).

\section{HST+JWST SED fitting}
\label{section:sedfitting}

	Having established the presence of outflows in at least some of these sources, we turn to the question of their star formation histories.  Studies of selected individual objects included in this sample with HST imaging suggest extremely young ages, low metallicities, compact sizes, and high specific star formation rates \citep{vanzella17}.  However, with the HST filters providing only coverage of the rest-frame UV (or, at best, optical for the lowest-redshift sources), it has been impossible to place strong constraints on the mass in any older stellar population.  With JWST/NIRCam imaging of two of the clusters included in the \cite{richard21} sample now available, providing up to rest-frame NIR coverage for sources with $ z \lesssim 4.0 $, we can now search for signatures of older stellar populations in these objects.

\begin{figure}[ht!]
	\centering
	\includegraphics[width=\textwidth]{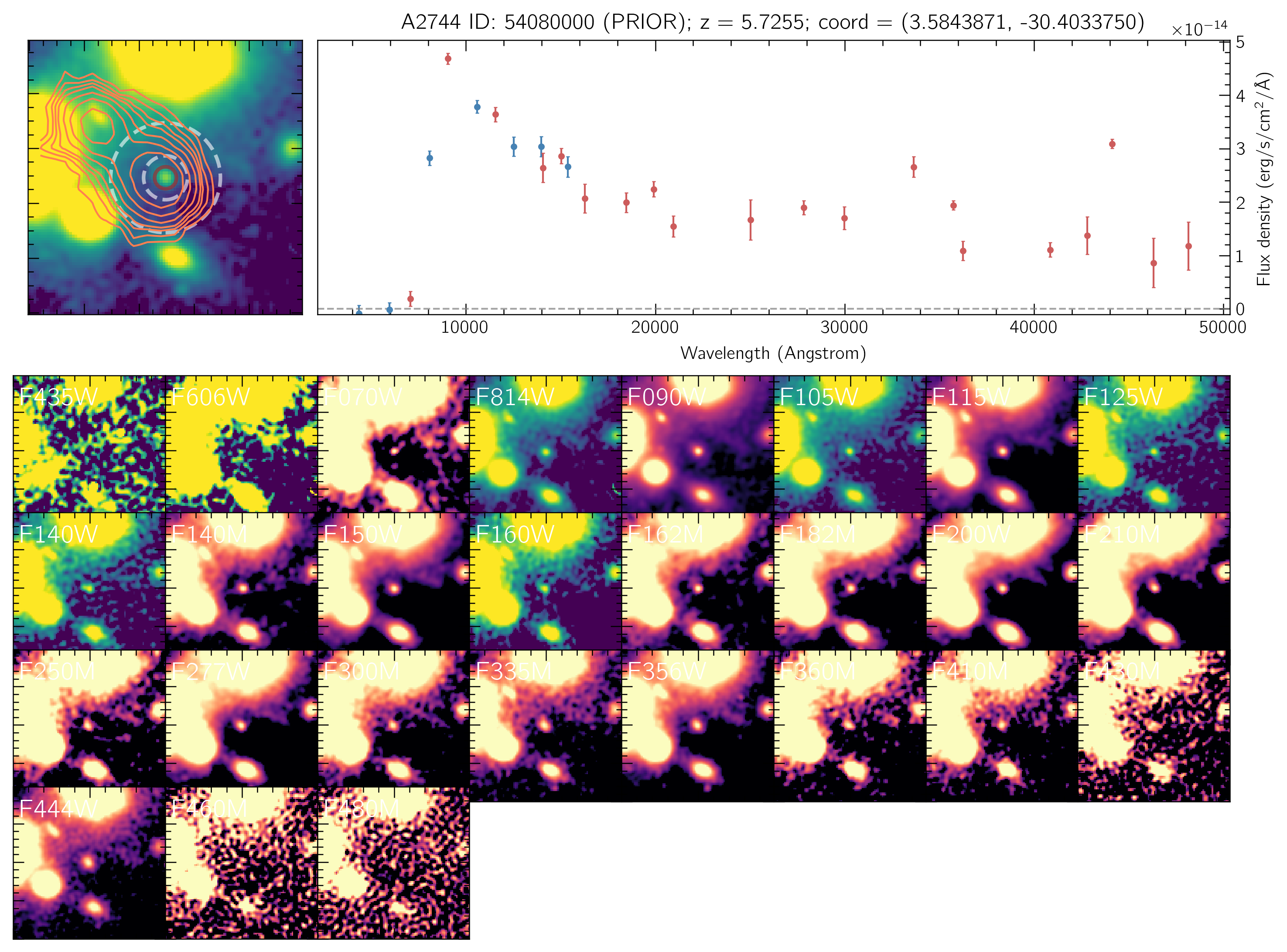}
	\caption{Example object from A2744.  Above: stacked image (left) with MUSE Lyman-$ \alpha $ contours (orange) and showing aperture with background annulus, and extracted SED (left) with HST measurements in blue and JWST in red. Below: individual postage stamp cutouts in each HST and JWST filter, arranges by wavelength.  The morphology appears consistent across all filters.}
	\label{fig:images}
\end{figure}
	
	We obtained JWST/NIRCam images of two clusters: MACS\,J0416.1-2403 (via the PEARLS collaboration; \citealt{windhorst23}) and Abell\,2744 (from Data Release 3 of the UNCOVER survey; \citealt{seuss24}), which we refer to as MACS0416 and A2744, respectively, for the remainder of this paper.  The PEARLS images of MACS0416 includes mostly broad-band and one medium-band filter (F090W, F115W, F200W, F277W, F356W, F410M, F444W), while the UNCOVER images of A2744 include these plus a number of additional medium-band filters (F140M, F162M, F182M, F210M, F250M, F300M, F335M, F360M, F430M, F460M, F480M) and one additional broad-band filter (F070W).  Both of these data sets are complemented by archival broad-band HST ACS and WFC images (F435w, F606W, F814W, F105W, F125W, F140w, F160W).  For the images of MACS0416, we aligned the images and matched the PSFs to the image with the coarsest angular resolution, which we found to be the HST/WFC F125W filter.  We then fitted light profiles to the bright, spectroscopically confirmed cluster members (using single Sersics, double Sersics, or core-Sersics), and subtracted them to mitigate contamination of the LAEs' photometry.  For the images of A2744, we simply obtained the aligned, PSF-matched, cluster-member-subtracted images directly from the UNCOVER data release\footnote{\url{https://jwst-uncover.github.io/DR3.html}}.
	
	As an initial check for older stellar populations, we visually compared the morphology of the sources in the HST and JWST images.  In many cases, we find no difference in morphology between the rest-frame UV and optical/NIR, as shown in the example in Figure \ref{fig:images}.  This demonstrates that, far from being star forming regions embedded in more massive galaxies, many of these objects are as compact as prior studies with HST suggest (e.g. \citealt{claeyssens22}).
	
	\begin{figure}[ht!]
		\includegraphics[width=\textwidth]{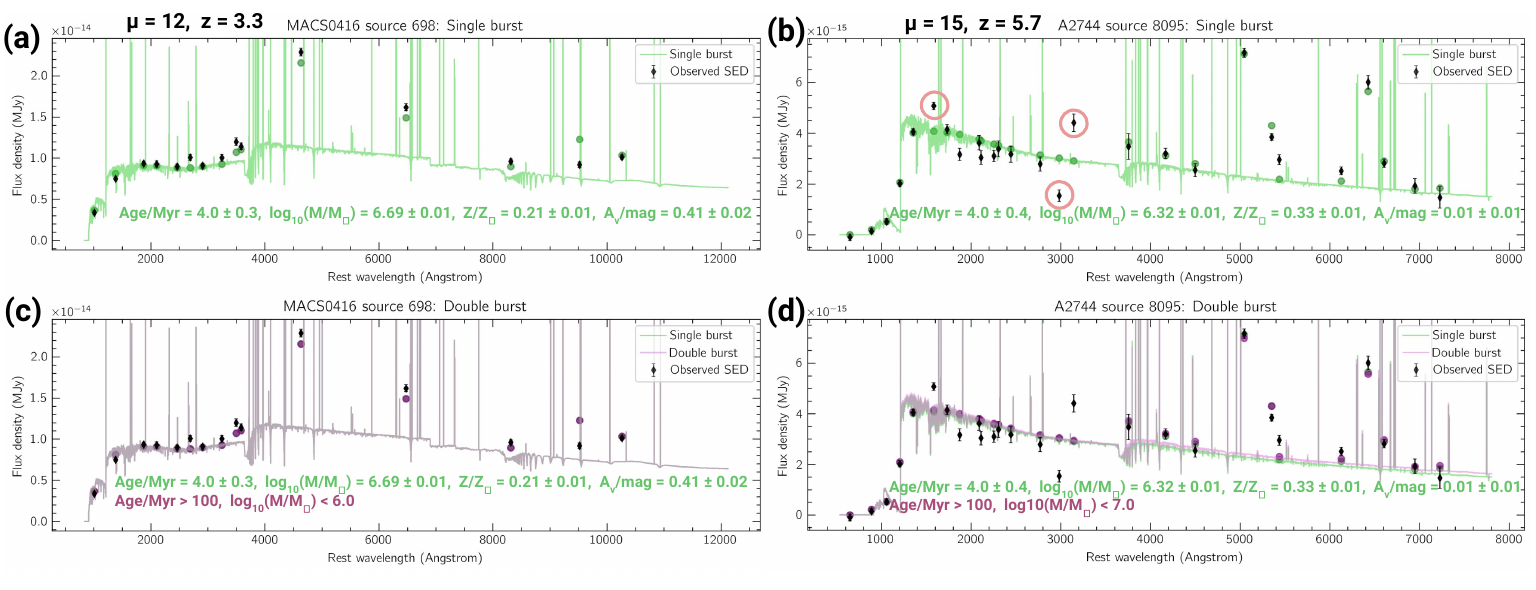}
		\caption{SEDs for two example objects with maximum likelihood models produced by \textsc{Bagpipes} for a single burst of star formation (panels (a) and (b)), and a young burst with a forced older burst (panels (c) and (d)).}
		\label{fig:sedfits}
	\end{figure}

	We performed apeture photometry on all the most highly magnified images of LAEs ($ \mu \geq 10 $), using magnifications and positions from the catalogues of \cite{richard21}.  In both clusters, we set the aperture diameter to be $ 2\times $ the full-width at half-maximum (FWHM) of the respective PSFs.  Having extracted SEDs for our objects, we then performed SED modelling using \textsc{Bagpipes} \citep{carnall18}.  We started by fitting the simplest possible star formation history: a single burst of star formation.  We find that in many, if not most, of the spectra, the maximum-likelihood single-burst model is already in good agreement with the observed SEDs as shown in Figure \ref{fig:sedfits}(a).  On the other hand, we find that some of the UNCOVER SEDs contain some points that are in very poor agreement with the model, by up to $ \sim 10\sigma $, as indicated with red circles in Figure \ref{fig:sedfits}(b).  Next, we try a star formation history consisting of two bursts of star formation, setting no age prior on one burst (the ``young" burst), and a uniform age prior $ > 100\,\text{Myr} $ on the other (the ``old" burst).  We show example two-burst models in Figure \ref{fig:sedfits} along with maximum likelihood parameters produced by \textsc{Bagpipes}.  In all cases, we find no improvement in the model fits by adding an old burst, and recover the same posterior distributions for the parameters of the young burst.  Consquently, we still find a few points in some of the the UNCOVER SEDs with large discrepencies between data and models.  These discrepencies appear largely random, and are not resolved by trying more complex star formation histories (exponential, constant, or power law).
	
	Our SED models show no evidence of older stellar populations in these objects.  However, inspecting the mass posteriors for the old bursts, we find a more complex story: while for some objects the mass of the old burst can be constrained to $ \sim $ an order of magnitude less than the young burst, as in Figure \ref{fig:sedfits}(c), in other cases the upper bounds are much less strict, and can even be higher than the mass in young stars as shown in Figure \ref{fig:sedfits}(d).  This demonstrates an important point: outshining by the young stellar population can be severe enough in these objects that even JWST/NIRCam observations are relatively insensitive to older stellar populations.  To convincingly rule out older generations of stars in these systems, either longer wavelength coverage with JWST/MIRI (e.g. \citealt{goovaerts23}), or else deeper observations with 30\,m-class telescopes are required.
	
\section{Acknowledgements}
\label{section:acknowl}

	J.L. and J.N. acknowledge support from the Research Grants Council of Hong Kong for conducting this work under the General Research Fund 17312122.  We also thank Rogier Windhorst and the rest of the PEARLS collaboration for providing us with early access to the JWST images of MACS0416.

\bibliographystyle{iaulike}
\bibliography{j_nianias_arxiv}

\end{document}